\documentclass[aps,pre,preprint,groupedaddress,showkeys]{revtex4}
\usepackage{epsfig}
\begin{document}

\preprint{}
\title{Bifurcation analysis of a model of the budding yeast cell cycle}


\author{Dorjsuren Battogtokh} 
\email{dbattogt@vt.edu, tel:  540-231-5508, fax:  540-231-9307}

\author{John J. Tyson} 
\affiliation{
Department of Biology, Virginia Polytechnic Institute and State University
Blacksburg, VA 24061
} 


\begin{abstract}
We study the bifurcations of a set of nine nonlinear ordinary differential 
equations that describe the regulation of the cyclin-dependent kinase that 
triggers DNA synthesis and mitosis in the budding yeast, 
{\em Saccharomyces cerevisiae}. We show that Clb2-dependent kinase 
exhibits bistability (stable steady states of high or low 
kinase activity). The transition from low to high Clb2-dependent
kinase activity is driven by transient activation of Cln2-dependent kinase,
and the reverse transition is driven by transient activation of the Clb2 
degradation machinery. We show that a four-variable model retains the main
features of the nine-variable model. In a three-variable
model  exhibiting birhythmicity (two
stable oscillatory states), we explore possible effects of extrinsic 
fluctuations on cell cycle progression.

\end{abstract}

\maketitle

\section{Introduction}
The cell cycle is the sequence of events by which a growing cell
replicates all its components and divides them evenly between two
daughter cells \cite{Watson, Nurse, Murray}. Many theoreticians
have understood the cell cycle as a periodic process driven by 
a biochemical limit cycle oscillator \cite{norel,obey,hatz}. 
However, a growing body of
experimental and theoretical evidences indicates that the eukaryotic
cell cycle is a toggle switch between two stable steady states,
controlled by checkpoints \cite{Nasmyth,Kathy,Cross,JTbook,Sible}.
This point of view was adopted by Chen et al. \cite{Kathy} in
a recent mathematical model of the budding yeast cell cycle, and 
bistability in the yeast cell cycle control system has been 
confirmed
recently by experiments from Cross's laboratory  \cite{Cross}. 

Bifurcation theory is a 
mathematical tool for characterizing  
steady state and oscillatory solutions of a system of 
nonlinear differential equations (ODE's) \cite{Kuznetsov,Strogatz}. 
The goal of this work is a detailed bifurcation analysis of 
Chen's model. Our  bifurcation analysis supplements the numerical simulation
carried out by Chen et al. and clarifies their quantitative comparisons 
between experiment and theory \cite{Cross}. In addition, bifurcation theory
helps us to identify control modules within Chen's complicated model,
thereby bringing some new insights to the yeast cell cycle control 
mechanism. A more through understanding 
of cell cycle control in yeast can be very helpful in future efforts to
model mammalian cell cycle controls \cite{kohn}. 

This paper is organized as follows. In Section II, we give a brief
introduction to the  budding yeast cell cycle.  In Section III, we 
introduce  Chen's model and present its one-parameter bifurcation diagram.
In Section IV, we study saddle node bifurcations in Chen's model in order 
to provide a rigorous foundation for interpreting Cross's experiment on 
bistability of the control system \cite{Cross}. In Section V, 
we propose  a reduced model with
four time-dependent variables, which retains the main dynamical characteristics 
of the extended model.  We characterize this model 
using two-parameter bifurcation diagrams. 
In Section VI, we  further reduce Chen's model to 
three variables and 
demonstrate that the abbreviated model displays bifurcations and birhythmicity 
similar to more complex  models {\cite{Borisuk,JTbook,Chaos}. 
In Section VII,  we use the three-variable model to study effects of 
extrinsic fluctuations.
The closing section is devoted to discussion. The nine-variable
mathematical model and its  parameters are given in the Appendix. 

\section{A Brief introduction to the budding yeast cell cycle}

{\em Cell cycle phases.} The cell cycle is the  process by 
which one cell  becomes two. The most important events in the
cell cycle are replication of the cell's DNA and separation of
the replicated DNA molecules to the daughter cells. In eukaryotic
cells, these events (replication and separation) occur in temporarily
distinct stages (S phase and M phase, respectively). S and M phase are 
separated in time by gaps called G1 and G2 phases.

During  S phase (``synthesis"), double-stranded DNA molecules are 
replicated to produce pairs of sister chromatids. During
M phase (``mitosis"), sister chromatids are separated 
so that each daughter cell receives a copy of each chromosome.
The G1  checkpoint mechanism controls the initiation of S phase,
and a  G2 checkpoint mechanism controls entry in  M phase.
A mitotic checkpoint controls the transition from M phase back to G1
phase. The checkpoints monitor cell size, DNA damage and repair,
DNA replication, and chromosome alignment on the mitotic spindle.

{\em Molecular controls of budding yeast cell cycle.}
Based on current knowledge 
about the molecular components controlling progression through the
budding yeast cell cycle, a molecular wiring diagram was proposed by 
Chen et al. \cite{Kathy}. A slightly simplified version of their
diagram is presented in  Figure 1. The molecular components can be divided
into four groups: cyclins, inhibitors, transcription factors, 
and proteolytic machinery. 

There are two families of cyclins in Figure 1: Cln's and Clb's
\footnote{we adopted following notations:
{\em ABC} denotes a gene, ABC implies a protein, and [ABC]
denotes a concentration for protein ABC.}. These
cyclins combine with kinase subunits (Cdc28) to form active 
cyclin-dependent kinase heterodimers that trigger cell cycle events
(Cdc28/Cln2 initiates budding, Cdc28/Clb5 initiates DNA synthesis, 
Cdc28/Clb2 initiates mitosis). 
Cdc28 subunits are  in constant, high abundance throughout cell 
cycle; hence, the activity of Cdc28/cyclin heterodimers is controlled
by the availability of cyclin subunits. For this reason, Cdc28 is not shown
in Figure 1; only the cyclin subunits are specified. (Each cyclin
molecule is understood to have a Cdc28 partner.)

Sic1 (in Figure 1) is a cyclin-dependent kinase inhibitor: it binds
to Cdc28/Clb dimers to form inactive trimers (Cdc28/Clb/Sic1). 
Sic1 does not bind to or inhibit Cdc28/Cln dimers.

Mcm1, MBF, SBF and Swi5 are transcription factors for synthesis of Clb2,
Clb5, Cln2 and Sic1, respectively.

The degradation of these proteins is regulated by a ubiquitination pathway. 
Proteins destined  for degradation  are first labeled by attachment of multiple
ubiquitin molecules. Ubiquitin  moieties are attached to Clb2 and Clb5 by the APC
(anaphase promoting complex) in conjunction with either Cdc20 or Hct1. 
Sic1 is ubiquitinated by a different mechanism(the ``SCF"), 
which (unlike the APC) requires that its substrates be phosphorylated.

Budding yeast cells progress through the division cycle as the levels of the species
in Figure 1 come and go. Thus the problem of cell cycle control is
to understand the temporal fluctuations  of these species. 
Because the species in Figure 1 are directly or indirectly 
interacting with all other species, simultaneous determination of their
fluctuating concentrations require a precise  mathematical model.
Using mass action and Michaelis-Menten rate laws, the complex wiring diagram in 
Figure 1 can be converted into ordinary differential equations, and
from them the molecular levels can be computed  \cite{Kathy}.

\section{A Bifurcation diagram of Chen's model}
The model proposed by Chen et al. \cite{Kathy} 
includes about a dozen ODE's
and  eleven algebraic equations with more than $50$ parameters. 
(Refer to  \cite{Kathy} for a complete description of 
the wiring diagram and a derivation of the  mathematical model, 
as well as for estimates of the rate constants in the model.)

In the appendix we present a reduced version of Chen's model
to be used in this paper for bifurcation analysis. From the original model,
we drop the target variables (spindle and bud formation, 
and DNA synthesis) because they are decoupled  from the rest
of the model. We  reserve  mass as the principal bifurcation parameter. 
We use the same parameter values as Ref. \cite{Kathy}, and they are presented
in the  appendix, Table I.

Using the  software package AUTO \cite {Doedel},
we created a one-parameter bifurcation diagram (Figure 2) 
of the budding yeast cell cycle model, Eqn. (A1-A20),
for parameter values given in Table I.  
Two saddle node bifurcations, at $M\approx 0.97$ and $M \approx 0.6$, 
connect the stable steady
states in Figure 2. There is also a subcritical Hopf bifurcation
on the upper branch of steady states at $M \approx 0.82$ from
which  a branch of unstable limit cycles originates. These unstable
oscillations disappear  at an infinite-period saddle loop (SL) bifurcation
near $M \approx 0.73$. A second branch of 
limit cycle oscillations,  shown by filled circles that 
disappear at a different  SL bifurcation point 
($M \approx 0.78$), are stable. 

The stable steady states (solid line) at values of [Clb2]$ < 10^{-3}$
represent the G1 phase of the cell cycle. The stable oscillatory states 
(filled circles) represent autonomous progression through S, G2 and
M phases of the cell cycle and then back into S phase. 

To get a full picture of cell cycle events, we must combine the 
dynamics of the cyclin-dependent kinase ``engine" (as summarized in Figure 2)
with equations for cell growth and division (changes in cell mass, $M$).
To this end, we supplemented Eqn. (A1-A20) with an equation for mass growth,
$M(t)=M(0) e^{\mu t}$, or, in differential form,  $\dot{M}=\mu M$, and a
rule for cell division ($M$ reset to $f M$ whenever Clb2-dependent kinase
activity drops below 0.005). Following Chen et al.  \cite{Kathy}
we choose $f=0.0043$ because budding yeast cells divide asymmetrically.

With these changes, we compute a solution of the full system, Eqn. (A1-A20)
plus the dynamics of $M$, and plot the resulting ``trajectory of motion" 
on the bifurcation diagram (the red line in Figure 2). 
This trajectory shows that 
the control system stays in the G1 phase if $M < 0.97$.
As  $M$ increases further, the control system is captured by the stable
limit cycle. As a result, ${\rm [Clb2]_T}$ increases abruptly,
driving the cell through S phase into M phase, then ${\rm [Clb2]_T}$
drops below 0.005, causing the cell to divide and the control system
to return to the stable G1 state. 

This  bifurcation diagram of the Chen et al. model exhibits the same
features of cell cycle models of frog eggs \cite{Borisuk}
and fission yeast \cite{Chaos}, namely, saddle node bifurcations
associated with stable and unstable oscillations. Yet,
there are subtle differences  in these bifurcation diagrams.
In the frog egg and fission yeast models, the large amplitude stable
limit cycles end at a saddle-node invariant-circle (SNIC) bifurcation,
not a SL bifurcation. In our case stable oscillations coexist
with the stable steady states over a small range of mass values
($0.78 <M<0.95$). However, when the budding yeast cell cycle
model is supplemented by the mass growth equation, such differences
seem unimportant. 


\section{Saddle-node bifurcations driven by Cln2 and Cdc20}

Recently, Cross et al. \cite{Cross} experimentally
confirmed bistablity in activity of Clb2-dependent kinase
in budding yeast cells. 
It is interesting to mention that this result was 
predicted by a  schematic sketch 
(Figure 9 of Ref. \cite{Kathy}) intuitively drawn from 
interrelations of Cdc28/Clb2 with the  
G1 phase cyclin Cln2 and the APC specificity factor Cdc20.
We confirm this informal  prediction of Chen et al.
by a rigorous  bifurcation analysis of their  model.

In their experimental work,  Cross et al. constructed a strain that under 
different experimental conditions may lack activities of 
Cln2 or Cdc20 or both. (To be precise, 
Cross et al. used  Cln3 in place of  Cln2,
but that technical detail makes no difference to our analysis.)
By manipulating  the activities of 
Cln2 or Cdc20, they found that [Clb2] can
be either in high or  low, depending on initial 
conditions. In terms of bifurcation theory, they provided evidence 
for an S shaped steady state curve bounded by saddle-node bifurcations,
with transitions driven by the activity of Cln2 or Cdc20.
Indeed, we found that Chen's model displays such
bifurcations when  [Cln2] and [Cdc20] are considered as
bifurcation parameters.

In accord with the experimental protocol of Cross et al. 
 \cite{Cross}, we consider [Cln2] and  [Cdc20] as parameters, 
and therefore discard
Eqn. (A1) and Eqn. (A4-A5) from Chen's model. We performed
bifurcation analysis for the remaining six ODE's. 
In Figure 3, we show a  combination of
two bifurcation diagrams.  In the left bifurcation diagram
we set [Cln2]=0 and vary [Cdc20], whereas
in the right bifurcation diagram we set [Cdc20]=0 and vary [Cln2]. 
As mass is the same in both cases ($M=1$), the two stable steady states in
Figure 3 represent G1 phase (low Clb2-dependent kinase activity)
and S/G2/M phase (high Clb2-dependent kinase activity).
Figure 3 shows that increasing 
[Cln2] drives the transition from  G1 into S/G2/M, 
while activation of [Cdc20] drives the transition from 
S/G2/M back to G1.

Using AUTO's facility for computing two-parameter bifurcation diagrams,
we extended the saddle-node bifurcations in Figure 3 into the parameter 
planes spanned by ([Cln2],$M$), ([Cdc20],$M$), and ([Cln2],[Cdc20]).
In Figure 4a, there are multiple steady states inside the cusp-shaped region 
bounded by the dashed lines, as expected \cite{Kuznetsov}. In Figure 4b, 
there are two different bistable domains, bounded by dashed and dotted lines, 
respectively. Where the domains overlap, we found that the control
 system has five steady states.  We found that two 
different modules independently lead to the bistable domains in Figure 4b.
The dashed line curve is due to the Hct1 module of the wiring 
diagram in Figure 1, 
whereas the dotted line curve is due to the Sic1  module.
Finally, Figure 4c shows the bistable region on the ([Cln2],[Cdc20]) plane.


\section{Effects of transcriptional factors Mcm1 and 
SBF in a reduced model with four ODE's}

Because Eqn. (A1-A20) take into account many known  details of cell 
cycle control, the model is very complex. It is difficult to understand 
from Eqn. (A1-A20) what are the nonlinearities leading to specific 
features of the bifurcation diagram shown in Figure 2. To 
overcome this difficulty, we simplify Chen's model, 
by defining a core module that retains the main dynamical 
features of the full set of equations. The 
reduced model can be useful in understanding the roles
of nonlinear feedbacks in the control system. 

In Figure 5 we propose a simplified wiring diagram for the  budding yeast
cell cycle. We discarded from the original wiring diagram the 
Sic1 and the Clb5 modules and Cdc20's activation, retaining only
four ODE's.

\begin{widetext}
\begin{eqnarray}
\frac{d}{dt} [{\rm Cln2}] = M (k_{s,n2}'+k_{s,n2}'' 
[{\rm SBF}]) - k_{d,n2} [{\rm Cln2}], 
\ \ \ \ \ \ \ \ \ \ \ \ \ \ \ \ \ \ \ \ \ \ \ \ \ \ \ \ \ \ \ \ \ \ \ \ \ \
\ \ \ \ \ \ \ \ \ \ \ \ \ \ \ \ \ \ \ \ \ \ \ \ \\ 
\frac{d}{dt}[{\rm Clb2}]=M (k_{s,b2}'+k_{s,b2}'' [{\rm Mcm1}] ) 
- ( k_{d,b2}'+(k_{d,b2}''-k_{d,b2}') [{\rm Hct1}]
+k_{d,b2}''' [{\rm Cdc20}]) [{\rm Clb2}],\ \ \ \ \ \  \ \ \ \ \ \\\
\frac{d}{dt}[{\rm Hct1}]=\frac{(k_{a,t1}'+k_{a,t1}'' 
[{\rm Cdc20}])(1-[{\rm Hct1}])}{J_{a,t1}+1-[{\rm Hct1}]}-\frac{V_{i,t1} 
[{\rm Hct1}]}{J_{i,t1}+[{\rm Hct1}]}, \: \ \ \  \ \ \ \ \ \  \ \ \
 \ \ \  \ \ \ \ \ \ \ \ \ \ \ \ \ \ \ \ \ \ \ \ \ \ \ \ \ \ \ \ \\
\frac{d}{dt}[{\rm Cdc20}]=(k_{s,20}'+k_{s,20}'' [{\rm Clb2}])
 -k_{d,20}' [{\rm Cdc20}],\ 
\ \ \ \ \ \ \ \ \ \ \ \ \ \ \ \ \ \ \ \ \ \ \ \ \ \ \ \ \ \ \ \ \ \ 
\ \ \ \ \ \ \ \ \ \ \ \ \ \ \ \ \ \ \ \ \ \ \ \ \   \; \\
\nonumber
\end{eqnarray}
\end{widetext}
where [SBF] is given by Eqn. (A13-A14) with [Clb5]=0. [Mcm1]
is given by Eqn. (A11), and  $V_{i,t1}$ is given by Eqn. (A15).
With the elimination of the dynamics for Cdc20 activation, 
we define a new parameter in Eqn. (4)
$k_{d,20}'=\frac{k_{d,20} k_{a,20}}{k_{a,20}+V_{i,20}+k_{d,20}}$. We note
that the results in this section do not change if $k_{d,20}'=k_{d,20}$.

Although Figure 5 is much simpler
than Figure 1, Eqn. (1-4) are still quite complex. The most uncertainties
arise from the two transcription factors(SBF and Mcm1), which are
described by nonlinear Goldbeter-Koshland functions 
\cite{Goldbeter,Koshland}. Their role
is to switch solutions from one branch to another. 
As the effects of the transcription factors
can be studied experimentally, 
we explore their  roles 
via two-parameter bifurcation diagrams.
First, by using mass  as the primary bifurcation parameter, we computed 
a one-parameter bifurcation diagram similar to
Figure 2. Then, we continued the codimension-one bifurcations
into two parameter domains, using 
the intensity coefficients of the transcription factors, $k_{s,n2}''$ 
and $k_{s,b2}''$, as the secondary bifurcation parameters. 

Figure 6a shows a two-parameter bifurcation diagram of Eqn. (1-4) on the
$(M,k_{s,n2}'')$ plane. Saddle-node bifurcations appear in Figure 6a
only if $k_{s,n2}''> 0$. A bistable domain  is inside the dashed 
lines.  It  widens at smaller mass and larger  $k_{s,n2}''$. 
The dot-dashed line showing 
Hopf bifurcation points continues inside the bistable domain. 
At $M >0.8$, the  Hopf bifurcation line is accompanied by a 
cyclic fold curve. These curves eventually 
coalesce at a larger mass value. Inside the bistable domain the cyclic fold
coalesces with a locus of saddle loops (solid line in Figure 6a).

Figure 6b shows a two-parameter bifurcation diagram on the $(M,k_{s,b2})$
plane. A crucial difference between Figure 6a and Figure 6b is
the existence of a bistable domain at $k_{s,b2}''=0$.  If
$k_{s,b2}''<0.04$, the effect of [Mcm1] regulation is negligible.
But if $k_{s,b2}''>0.5$,  [Mcm1] can destroy bistability. In Figure 6b
a Hopf bifurcation line originates from a Bogdanov-Takens 
bifurcation. This line  is accompanied by a  line of saddle loops. 
The saddle loops 
change stability where the line of cyclic folds coalesces with the line of
saddle loops.

We found that   Eqn. (A1-A20) display  two-parameter 
bifurcation diagrams similar to Figure 6a-b. 
Notice from Figure 6b that the domain of bistability is quite
independent of the activity of Mcm1, but the existence of the primary
Hopf bifurcation in the model is sensitively dependent on the activity of Mcm1. 

\section{A SNIC bifurcation in
a reduced model with three ODE's}

The eukaryotic 
cell cycle engine is a highly conserved molecular machine.
It is expected that mathematical models of cell cycle controls in 
different organisms exhibit qualitatively similar dynamics as revealed by
similar bifurcation diagrams. But there can be also peculiarities in 
these models, subject to particular parameter selections.
As we mentioned in Section 3, the bifurcation diagram in Figure 2 
does not involve a SNIC bifurcation, as seen in 
 bifurcation diagrams of mathematical models for 
frog eggs and fission yeast \cite{Borisuk,Chaos,Hunding}.
Although this difference is  rather subtle 
and does not contradict any features of  cell cycle physiology, 
we point out  that Chen's model Eqn. (A1-A20)  
can display a SNIC bifurcation for appropriate choice of parameter
values (not shown). In this section, we examine SNIC bifurcation in a three
variable model. 

To further simplify the model, we neglect Cln2 from 
the wiring diagram in Figure 5.
As a result, we have a model with three time-dependent variables,

\begin{widetext}
\begin{eqnarray}
\frac{d}{dt}[{\rm Clb2}]=M (k_{s,b2}'+k_{s,b2}'' [{\rm Mcm1}] ) 
- ( k_{d,b2}'+(k_{d,b2}''-k_{d,b2}') [{\rm Hct1}]
+k_{d,b2}''' [{\rm Cdc20}]) [{\rm Clb2}],\ \ \ \ \ \  \ \ \ \ \ \\\
\frac{d}{dt}[{\rm Hct1}]=\frac{(k_{a,t1}'+k_{a,t1}'' 
[{\rm Cdc20}])(1-[{\rm Hct1}])}{J_{a,t1}+1-[{\rm Hct1}]}-\frac{V_{i,t1} 
[{\rm Hct1}]}{J_{i,t1}+[{\rm Hct1}]}, \: \ \ \  \ \ \ \ \ \  \ \ \
 \ \ \  \ \ \ \ \ \ \ \ \ \ \ \ \ \ \ \ \ \ \ \ \ \ \ \ \ \ \ \ \\
\frac{d}{dt}[{\rm Cdc20}]=(k_{s,20}'+k_{s,20}'' [{\rm Clb2}])
 -k_{d,20} [{\rm Cdc20}].\ 
\ \ \ \ \ \ \ \ \ \ \ \ \ \ \ \ \ \ \ \ \ \ \ \ \ \ \ \ \ \ \ \ \ \ 
\ \ \ \ \ \ \ \ \ \ \ \ \ \ \ \ \ \ \ \ \ \ \ \ \   \; \\
\nonumber
\end{eqnarray}
\end{widetext}

In Eqn.(5-7), [Mcm1] is given by Eqn. (A11), and  
$V_{i,t1}$ is given by Eqn. (A15). We assume
${\rm [Cln2]=0}$ and ${\rm [Clb5]} =\frac{k_{s,b5}' M}{k_{d,b5}'}$ in Eqn. (A15).
We also changed the values of some parameters in Table I, as $k_{s,b5}'=0.06$, 
$k_{a,t1}''=1.5$, $k_{s,20}''=0.07$, $J_{a,mcm}=J_{i,mcm}=0.01$. 

Let $[{\rm Clb2} ]^0$, $[{\rm Hct1}]^0$ and ${[{\rm Cdc20}]}^0$
denote a steady state solution of Eqn. (5-7). 
Clearly, ${[{\rm Cdc20}]}^0=\frac{(k_{s,20}'+k_{s,20}'' 
{[{\rm Clb2}]}^0)}{k_{d,20}}$. Substituting this functional relation
between [Cdc20] and [Clb2] into Eqn. (5-6), we can think of 
([Clb2],[Hct1]) as a two-variable system, susceptible to phase plane
analysis. The nullclines of the two variable systems are plotted in 
Figure 7. From the intersections of these nullclines, we find steady
state solutions,  $[{\rm Clb2} ]^0$ and $[{\rm Hct1}]^0$, and
consequently ${[{\rm Cdc20}]}^0$.  Depending on 
$M$, the number of intersections varies, but the 
maximum number of steady states is three. 

We study stability of the steady states numerically. In Figure 8
we plot  a bifurcation diagram of Eqn. (5-7), with $M$
as the principal bifurcation parameter. At 
 a given $M$, there can be one stable steady state and
two unstable steady states, or a single steady state which 
can be either stable or unstable. The stable steady states
in Figure 8 can coexist 
with stable limit cycle oscillations.  There are two 
interesting features in this bifurcation diagram: (i)
a SNIC bifurcation which arises at $M\approx 2.9$ 
where a saddle-node  coalescence 
is replaced by limit cycle oscillations, and (ii)
byrhithmicity, i.e., coexistence of two stable limit cycle oscillations,
for $2.8<M<4.9$. 

Figure 9 shows a two-parameter bifurcation diagram for Eqn. (5-7). Despite 
the reduction to just three ODE's, this
diagram is  quite complex. Multiple steady states 
are found inside the solid lines. There are three Bogdanov-Takens 
bifurcations in Figure 9, from which originate three independent loci of
Hopf bifurcations, shown by  lines in violet. Two cyclic folds associated
with the Hopf bifurcations are shown by 
lines in cyan. Two saddle loops, shown by green lines,
originate at ${\rm BT_1}$ and $\rm{ BT_2} $, cross the region of
bistability, and attach to the right saddle node line at two
saddle-node-loop bifurcation points. 
Between these two saddle-node-loops,  we find 
a SNIC bifurcation, red line in  Figure 9. 

\section{Birhythmicity and effects of extrinsic fluctuations}

Figure 9 shows that
the distance between 
the inner cyclic fold(${\rm CF_2}$) and the SNIC bifurcation varies as
$k_{s,b2}''$ changes. 
In other words,  depending on $k_{s,b2}''$, birhythmicity may occur 
either close to, or far  from the START transition (when the stable
G1 state gives way to large amplitude stable oscillations). If it happens 
far away from START, it will not interfere with cell cycle progression.
However, if it occurs close to START, as in Figure 8,
an interesting question arises.  To which 
stable oscillation (the large amplitude or the small amplitude limit
cycles) 
will the trajectory of motion (see red lines in Figure 2) connect?
We found that  the trajectory of motion always
follows the large amplitude 
slow oscillations in the three-variable model.
We have shown (in a separate publication) 
 that switching between small and large
amplitude oscillations is possible when 
the model takes into account diffusion terms \cite{battyson}.
Here, we demonstrate the effects of 
noise on the trajectory of motion.

A complex process, such as cell cycle control, is naturally
subject to fluctuations from different sources.
For instance, stochastic effects due to 
size and nuclear volume differences at cell 
division have been studied for fission yeast \cite{Sveiczer}. 
Since we know very little about the origin of fluctuations
in the cell cycle engine, the simplest way to incorporate random processes 
into Eqn. (5-7) is to assume  that certain 
extrinsic fluctuations randomly 
perturb the cell cycle engine.
Mathematically, we replace  Eqn. (5-7)  by  Langevin-type equations 
with multiplicative noise \cite{Kampen,steuer},

\begin{eqnarray}
\frac{d}{dt}[{\rm Clb2}]=F_{[{\rm Clb2}]}+\sqrt{2 D_1 [{\rm Clb2}]} \xi(t),
\ \ \ \ \ \ \ \ \ \ \ \ \ \ \ \ \ \ \ \ \ \ \ \ \ \ \ \ \ \ 
\ \ \ \ \ \ \ \ \ \ \ \ \ \ \ \ \ \ \ \ \ \ \ \ \ \ \ \ \ \ \\
\frac{d}{dt}[{\rm Hct1}]=F_{[{\rm Hct1}]}+\sqrt{2 D_2 [{\rm Hct1}]}  \xi(t), 
\ \ \ \ \ \ \ \ \ \ \ \ \ \ \ \ \ \ \ \ \ \ \ \ \ \ \ \ \ \ 
\ \ \ \ \ \ \: \ \ \ \ \ \ \ \ \ \ \ \ \ \ \ \ \ \ \ \ \ \ \\
\frac{d}{dt}[{\rm Cdc20}]=F_{[{\rm Cdc20}]}+\sqrt{2 D_3 [{\rm Cdc20}]}  \xi(t),
\ \ \ \ \ \ \ \ \ \ \ \ \ \ \ \ \ \ \ \ \ \ \ \ \ \ \ \ \ \ 
\ \ \ \ \ \ \ \ \ \ \ \ \ \ \ \ \ \ \ \ \ \ \ \ \\
\frac{dM}{dt}=\mu M.
\ \ \ \ \ \ \ \ \ \ \ \ \ \ \ \ \ \ \ \ \ \ \ \ \ \ \ \ \ \ 
\ \ \ \ \ \ \ \ \ \ \ \ \ \ \ \ \ \ \ \ \ \ \ \ \
\ \ \ \ \ \ \ \ \ \ \ \ \ \ \ \ \ \ \ \ \ \ \ \ \
\ \ \ \ \ \ \ \ \ \ \ \: \:
 \\
\nonumber
\end{eqnarray}
where, $F_{[{\rm Clb2}]}$, $F_{[{\rm Hct1}]}$, $F_{[{\rm Cdc20}]}$ are the
right hand sides of Eqn. (5-7) and 
$\xi(t)$ is  Gaussian white noise with zero 
mean and unit variance,
\begin{equation}
<\xi(t)=0>, \ \ \ \ \ \ \ \ \ \ <\xi(t) \xi(t')>=\delta(t-t').
\end{equation}

We assume that mass increase is not affected by  random  fluctuations
\cite{steuer}.

We simulated Eqn. (8-11) using standard numerical techniques for
stochastic differential equations \cite{Sancho1,Sancho2}. In 
Figure 10 we overplot two different simulations. The dashed lines
show time evolutions  of $M$, [Clb2], [Cdc20] and  [Hct1]
when birhythmicity occurs far from START.
In this case, noise does not interfere with 
mitosis, and cell mass divides each time  [Clb2]
drops below $0.1$. The solid lines show the case when
birhythmicity occurs close to  START, as in Figure 8. In this
case noise can switch the control system from slow, 
large amplitude oscillations to fast, small amplitude oscillations. 
As a result, [Clb2] does not go below $0.1$ and the cell 
cannot divide. Consequently, mass $M$ grows and the
system goes to the stable steady state (see filled 
diamonds at $M>3.9$ in Figure 8). 
Therefore, in the presence of noise,  
birhythmicity may lead to mitotic arrest. 
\section{Discussion}
In this work, we carried out bifurcation analysis of a model of 
the budding yeast cell cycle, based on earlier work by Chen et al.
\cite{Kathy}  which successfully accounts for many observed features of 
proliferating yeast cells. Our results show that, 
despite a peculiarity in  topology of 
the bifurcation diagram, the budding yeast cell cycle model 
displays the same basic  features previously associated with 
frog egg and fission yeast models;  namely, 
saddle-node bifurcations associated with stable and unstable oscillations.

We explored  bistability and hysteresis in this model by
numerical bifurcation analysis. 
Some of our bifurcation diagrams can be useful for designing new
experiments. For instance, our two parameter bifurcation analysis
(Figure 4b) suggests  that the [Hct1] and [Sic1] modules may lead
independently to bistable states, and there can be regions in 
parameter space  with three stable steady states,
when these two modules operate cooperatively. 

We found that a reduced model with four time-dependent
variables retains the main characteristics of the bifurcation diagram
of Chen's model. This reduction allows us
to explore  the dominant roles of SBF and Mcm1 transcription
factors in budding yeast checkpoint controls.
Our two-parameter bifurcation diagrams (Figure 6)  also
can be useful in designing experiments for cell cycle controls by  
transcription factors.

The  budding yeast cell cycle model of Chen et al. is parameter rich.
Although the parameter set presented in Table I
leads to a satisfactory fit of the model to many experimental
observations, 
the choice of parameter values should be further constrained 
by new biochemical data about the protein-protein interactions
and further improved by automatic 
parameter estimation techniques \cite{jason,bat}. 
On the other hand, different sets of parameters, leading 
to different bifurcation scenarios, are interesting  
from a theoretical standpoint. We have proposed a set of parameters for
a reduced, three-variable model leading to a SNIC bifurcation.
 
An interesting feature accompanying the appearance
of a SNIC bifurcation in the reduced model 
is birhythmicity. Birhythmicity
has been found in a chemical system \cite{Birythm}, but 
for biological systems, it is known theoretically only 
\cite{Goldbeter,Moran,Borisuk}.
We have shown that in the presence of extrinsic fluctuations,
birhythmicity can lead to mitotic arrest. 
The fact that noise can switch a biochemical system from
one stable solution to another is well known (e.g. Ref. 
\cite{adams,Hasty}), but switching from one stable oscillations
to another is a less studied research area. A more systematic study 
of switching between stable limit cycles is a problem for the future.

\acknowledgments
Authors thank Kathy Chen and other members of Computational
Cell Biology Group at Virginia Tech for many 
stimulating discussions. This work was supported by a grant from DARPA's 
Biocomputation Program(AFRL $\#$F300602-02-0572). 

\newpage

\centerline{\bf Figure Captions}

Fig. 1. Wiring diagram of a budding yeast cell cycle model \cite{Kathy}.

Fig. 2. A one-parameter bifurcation diagram of Eqn. (A1-A20)
for parameter values in Table I.
Solid lines indicate stable steady states. Dashed lines indicate 
unstable steady states. Solid circles denote the maximum and minimum
values of  $[{\rm Clb2}]_T$ on 
stable limit cycle oscillations, open circles denote the same
for  unstable oscillations. The 
red line shows the trajectory of motion when Eqn. (A1-A20) are
supplemented by the mass growth equation $\dot{M}=\mu M$. The cell
divides ($M=f \cdot M$) when $[{\rm Clb2}]_T$ drops below $0.005$.

Fig. 3. Bistability and hysteresis driven by [Cln2] and [Cdc20].
On the left plane $[{\rm Cln2}] \equiv 0$, on the right plane 
$[{\rm Cdc20}] \equiv 0$.
Mass is fixed at 1. Filled diamonds show stable steady states,
dashed lines show unstable steady states.
Dotted lines and arrows indicate  the START and FINISH 
transitions of  the hysteresis loop. (START refers to the 
G1 $\longrightarrow$ S
transition, FINISH refers to the M  $\longrightarrow$ G1 transition.)

Fig. 4a. Two-parameter bifurcation diagram on the $([{\rm Cln2}],M)$
plane. Multiple steady states are found inside the cusp-shaped curve.

Fig. 4b. Two-parameter bifurcation 
diagram on the $([{\rm Cdc20]},M)$ plane. There are two independent 
pairs of saddle-node bifurcation curves in this figure
(dashed curves and dotted curves).
Depending on the overlaps of the regions bounded by these curves, the 
number of steady states varies from I to V.

Fig. 4c. Two parameter bifurcation diagram on the $([{\rm Cln2}],
[{\rm Cdc20}])$ plane. 
Bistable steady states are found in between the  dashed curves. 

Fig. 5. Wiring diagram of a reduced model with four-variables. 

Fig. 6a. Two-parameter bifurcation diagram of Eqn. (1-4) on the $(k_{s,n2}'',M)$
plane. Loci of saddle node bifurcations (SN) are shown by
dashed lines. Other lines trace saddle loop (solid), 
cyclic fold
(dotted), and Hopf (dot-dash) bifurcation points.

Fig. 6b. Two-parameter bifurcation diagram 
of Eqn. (1-4) on the $(k_{s,b2}'',M)$ plane. Bistable steady states 
are found inside the SN curve (dashed line). 
Other lines indicate loci of saddle loops (solid),
cyclic folds (dotted), and Hopf bifurcations (dot-dash).
The locus of Hopf bifurcations 
originates from a Bogdanov-Takens point shown by the filled circle.

Fig. 7. Stationary solutions of Eqn. (5-6) can be
computed from  the intersections of the [Hct1] nullcline  (solid line)
and the  [Clb2] nullcline (dashed line). In this plot mass is fixed at $M=2$.
Notice that $\rm{[Clb2]} \approx 0.15$ is the region where 
[Mcm1] changes abruptly from 0 to 1. 

Fig. 8. Bifurcation diagram of Eqn. (5-7). Filled diamonds 
show stable steady states, dashed lines show unstable steady states.
Stable limit cycle oscillations are shown by filled circles,
unstable limit cycle oscillations are shown by open
circles.

Fig. 9. Two-parameter bifurcation diagram of Eqn. (5-7) on the
$(M,k_{s,b2}'')$ plane. Bistability is found inside the SN curve (solid line).
Three different Hopf bifurcations (violet lines) originate from three 
Bogdanov-Takens bifurcation points shown by filled circles at 
${\rm BT_1}$, $\rm{ BT_2}$ 
and $\rm {BT_3}$. Cyclic folds are shown by lines in cyan, saddle loops 
by lines in green, and the red solid line shows  SNIC bifurcations.
${\rm SL_3}$, which runs next to ${\rm HB_3}$, is not shown on the diagram.
${\rm CF_2}$ runs from a degenerate Hopf bifurcation on ${\rm HB_1}$
to a degenerate Hopf bifurcation on ${\rm HB_2}$. ${\rm CF_1}$  runs from
a degenerate saddle loops on ${\rm SL_1}$  to a degenerate saddle loop on 
${\rm SL_2}$ (not shown), crossing over ${\rm HB_1}$
on the way. Where ${\rm CF_1}$ and ${\rm CF_2}$ run very close together, 
only ${\rm CF_2}$ is plotted on the figure. 

Fig. 10. Stochastic simulations of Eqn. (8-11).
Dashed lines show a case when birhythmicity
occurs far  from the SNIC bifurcation. In this case,  noise does not
interfere with cell cycle progression. Solid lines show the case
when birhythmicity occurs close to  the SNIC bifurcation, 
as in Figure 8. In the presence of noise, the latter case leads 
eventually to 
mitotic arrest. Parameters are: $D_1=D_2=D_3=3.75\cdot10^{-5}$. 
Solid lines for $k_{s,b2}''=0.05$,
dashed lines for $k_{s,b2}''=0.06$.

\newpage
\appendix
\section{Model for budding yeast cell cycle}

\begin{widetext}
\begin{eqnarray}
\frac{d}{dt} [{\rm Cln2}] = M (k_{s,n2}'+k_{s,n2}'' [{\rm SBF}]) - 
k_{d,n2} [{\rm Cln2}], 
\ \ \ 
\ \ \ \ \ \ \ \ \ \ \ \ \ \ \ \ \ \ \ \ \ \ \ \ \ \ \ \ \ \ \ \ \ \ \ \ \ \
\ \ \ \ \ \ \ \ \ \ \ \ \ \ \ \ \ \ \ \ \ \ \ \ \\ 
\frac{d}{dt}[{\rm Clb2}]_T=M (k_{s,b2}'+k_{s,b2}'' [{\rm Mcm1}] ) 
- ( k_{d,b2}'+(k_{d,b2}''-k_{d,b2}') [{\rm Hct1}]
+k_{d,b2}''' [{\rm Cdc20}]) [{\rm Clb2}]_T,\ \ \ \ \ \ \; \ \ \ \ \\\
\frac{d}{dt}[{\rm Hct1}]=\frac{(k_{a,t1}'+k_{a,t1}'' 
[{\rm Cdc20}])(1-[{\rm Hct1}])}{J_{a,t1}+1-[{\rm Hct1}]}-\frac{V_{i,t1} 
[{\rm Hct1}]}{J_{i,t1}+[{\rm Hct1}]}, \: \ \ \  \ \ \ \ \ \  \ \ \
 \ \ \  \ \ \ \ \ \ \ \ \ \ \ \ \ \ \ \ \ \ \ \ \ \ \ \ \ \ \ \ \ \ \ \\
\frac{d}{dt}[{\rm Cdc20}]_T=(k_{s,20}'+k_{s,20}''
 [{\rm Clb2}]) -k_{d,20} [{\rm Cdc20}]_T,\ 
\ \ \ \ \ \ \ \ \ \ \ \ \ \ \ \ \ \ \ \ \ \ \ \ \ \ \ \ \ \ \ \
\ \ \ \ \ \  \ \ \ \ \ \ \ \ \ \ \ \ \ \ \ \ \ \ \ \ \ \; \\
\frac{d}{dt}[{\rm Cdc20}]=k_{a,20}([{\rm Cdc20}]_T-[{\rm Cdc20}])-(V_{i,20}+k_{d,20})
[{\rm Cdc20}],  \ \ \ \ \ \ \ \ \ \ \ \ \ \ \ \ \ \ \ \ \ \ \ \ \ \ \ \ \ \ 
 \: \ \ \ \ \ \ \ \ \ \ \  \ \ \ \\
\frac{d}{dt} [{\rm Clb5}]_T = M (k_{s,b5}'+k_{s,b5}'' [{\rm MBF}]) - 
(k_{d,b5}'+k_{d,b5}''[{\rm Cdc20}]) [{\rm Clb5}]_T, \; \
\ \ \ \ \ \ \ \ \ \ \ \ \ \ \ \ \ \ \ \ \ \ \ \ \ \ \ \ \ \ \ \ \ \ \ \ \ \
\\ 
\frac{d } {dt}[{\rm Sic1}]_T= k_{s,c1}'+k_{s,c1}'' [{\rm Swi5}]
 - (k_{d1,c1}+\frac{V_{d2c1}}{J_{d2,c1}+[{\rm Sic1}]_T})
[{\rm Sic1}]_T,\ \ \ \ \ \ \ \ \ \ \ \ \ \ \ \ \ \ \ \ \ \ \ \ \ \ \ \ \ \:
\ \ \ \ \ \ \ \ \ \ \ \ \ \\
\frac{d} {dt} [{\rm Clb5|Sic1}]= k_{as,b5} [{\rm Clb5}] [{\rm Sic1}] -\nonumber
\ \ \ \ \ \ \ \ \ \ \ \ \ \ \ \ \ \ \ \ \ \ \ \ \ \ \ \ \ \ \ \ \ \ \:
\ \ \ \ \ \ \ \ \ \ \ \ \ \ \ \ \ \ \ \ \ \ \ \ \ \ \ \ \ \ \ \
\ \ \ \ \ \ \ \ \ \ \ \ \ \ \ \! \!  \\
- (k_{di,b5}+ k_{d,b5}'+ k_{d,b5}'' [{\rm Cdc20}]
+k_{d1,c1}+\frac{V_{d2c1}}{J_{d2,c1}+[{\rm Sic1}]_T}) [{\rm Clb5|Sic1}], 
 \ \  \ \ \ \ \ \ \ \ \ \ \ \ \ \ \ \ \ \ \
\ \ \ \ \ \  \ \  \ \ \ \ \ \ \ \ \\
\frac{d} {dt} [{\rm Clb2|Sic1}]= k_{as,b2} [{\rm Clb2}] [{\rm Sic1}] -\nonumber
\ \ \ \ \ \ \ \ \ \ \ \ \ \ \ \ \ \ \ \ \ \ \ \ \ \ \ \ \ \ \ \ \:
\ \ \ \ \ \ \ \ \ \ \ \ \ \ \ \ \ \ \ \ \ \ \ \ \ \ \ \ \ \ \ \:
\ \ \ \ \ \ \ \ \ \ \ \ \ \ \ \ \\
- (k_{di,b2}+ 
( k_{d,b2}'+(k_{d,b2}''-k_{d,b2}') [{\rm Hct1}]
+k_{d,b2}''' [{\rm Cdc20}])
+k_{d1,c1}+\frac{V_{d2c1}}{J_{d2,c1}+[{\rm Sic1}]_T}) [{\rm Clb2|Sic1}],
\ \ \ \ \ \ \\
V_{d2,c1}=k_{d2,c1}(\epsilon_{c1,n3} [{\rm Cln3}]^*+\epsilon_{c1,k2} [{\rm Bck2}]+[{\rm Cln2}]+
\epsilon_{c1,b5} [{\rm Clb5}]+\epsilon_{c1,b2} [{\rm Clb2}]), \ \ \!
\ \ \ \ \ \ \ \ \  \ \ \:  \ \ \ \ \ \  \\
{[{\rm Mcm1}]} =G(k_{a,mcm}[{\rm Clb2}],k_{i,mcm}, J_{a,mcm},J_{i,mcm}) ,
\ \ \ \ \ \ \ \ \ \ \  \ \ \ \ \ \ \ \ \ \ \ \  \ \ \ \ \; \ \ \ \ 
 \ \ \ \ \ \ \ \ \ \ \ \ \ \ \ \ \ \ \ \ \  \ \ \ \ \\
{[{\rm Swi5}]} =G(k_{a,swi}[{\rm Cdc20}],k_{i,swi}'+k_{i,swi}'' [{\rm Clb2}]
, J_{a,swi},J_{i,swi}),\ \!
\ \ \ \ \ \ \ \ \ \ \  \ \ \ \ \ \ \ \  \ \ \ \ 
 \ \ \ \ \ \ \ \ \ \ \ \ \ \ \ \ \ \ \ \ \ \\
{[{\rm SBF}]}={[{\rm MBF}]}=G(V_{a,sbf},k_{i,sbf}'
+k_{i,sbf}''[{\rm Clb2}],J_{a,sbf},J_{i,sbf}), \!
\! \ \ \ \ \ \ \ \ \ \  \ \ \ \ \ \ \ \ \ \ \ \  \ \ \ \ \ \ \ \
\ \  \ \ \ \ \ \ \ \ \ \ \ \\
V_{a,sbf}=k_{a,sbf}([{\rm Cln2}]+\epsilon_{sbf,n3} ( [{\rm Cln3}]^*+ [{\rm Bck2}])
+\epsilon_{sbf,b5} [{\rm Clb5}]), \ \  \ \ \ \ \ \ \ \ \ 
\ \  \ \ \ \ \ \ \ \ \ \ \ \ \
\ \  \ \ \ \ \ \ \ \ \ \ \\
V_{i,t1}=k_{i,t1}'+k_{i,t1}''([{\rm Cln3}]^*+\epsilon_{i,t1,n2} 
[{\rm Cln2}]+\epsilon_{i,t1,b5} [{\rm Clb5}]+\epsilon_{i,t1,b2} [{\rm Clb2}]), \
\ \ \ \ \ \ \ \ \  \ \ \ \ \ \ \ \ \ \ \ \ \ \ \ \ \ \ \\
{[{\rm Clb2}]_T}=[{\rm Clb2}]+[{\rm Clb2|Sic1}],
\ \ \ \ \ \ \ \
\ \ \ \ \ \ \ \ \ \ \ \ \ \ \ 
\ \ \ \ \ \ \ \ \ \ \ \ \ \ \ 
\ \ \ \ \ \ \ \ \ \ \ \ \ \ \
\ \ \ \ \ \ \ \ \ \ \ \ \ \ \ \ \ \ \ \ \ \ \ \ \ \ \ \\
{[{\rm Clb5}]_T}=[{\rm Clb5}]+[{\rm Clb5|Sic1}],
\ \ \ \ \ \ \ \
\ \ \ \ \ \ \ \ \ \ \ \ \ \ \ 
\ \ \ \ \ \ \ \ \ \ \ \ \ \ \ 
\ \ \ \ \ \ \ \ \ \ \ \ \ \ \
\ \ \ \ \ \ \ \ \ \ \ \ \ \ \ \ \ \ \ \ \ \ \ \ \ \ \ \\
{[{\rm Sic1}]_T}=[{\rm Sic1}]+[{\rm Clb2|Sic1}]+[{\rm Clb5|Sic1}],\!\!
\ \ \ \ \ \ \
\ \ \ \ \ \ \ \ \ \ \ \ \ \ \ 
\ \ \ \ \ \ \ \ \ \ \ \ \ \ \
\ \ \ \ \ \ \ \ \ \ \ \ \ \ \ \ \ \ \ \ \ \ \ \ \ \ \ \\
{[{\rm Bck2}]}=M [{\rm Bck2}]^0,\!\
\ \ \ \ \ \ \ \ \ \ \ \ \ \ \ \ \ \ \ \ \
\ \ \ \ \ \ \ \ \ \ \ \ \ \ \ 
\ \ \ \ \ \ \ \ \ \ \ \ \ \ \ 
\ \ \ \ \ \ \ \ \ \ \ \ \ \ \
\ \ \ \ \ \ \ \ \ \ \ \ \ \ \ \ \ \ \ \ \ \ \ \ \ \ \ \\
{[{\rm Cln3}]^*}=[{\rm Cln3}]_{max} \frac{M D_{n3}}{J_{n3} + M D_{n3}}. 
\ \ \ \ \ \ \ \ \ \ \ \ \ \ \ \ \ \ \
\ \ \ \ \ \ \ \ \ \ \ \ \ \ \ 
\ \ \ \ \ \ \ \ \ \ \ \ \ \ \ \:
\ \ \ \ \ \ \ \ \ \ \ \ \ \ \ \ \ \ \ \ \ \ \ \ \ \ \ \\
\nonumber
\end{eqnarray}
\end{widetext}
Golldbeter Koshland function:
\begin{equation}
G(a,b,c,d)= \frac{2 a d}{ b-a+b c +a d +\sqrt{(b-a+b c +a d)^2-4 a b (b-a)}}
\end{equation}


\begin{table}
\caption{\label{tab:table2}
Kinetik constants for the budding yeast model}

\begin{ruledtabular}
\begin{tabular}{lllllll}
\hline
Rate constants($min^{-1}$)\\
$k_{s,b5}'=0.006$
  \ \ \ \ \ \ \ \ \ $k_{s,b5}''=0.02$  \ \ \ \ \ \ \ \
$k_{d,b5}'=0.1$ \ \ \ \ \ \ \ \ \ \ $k_{d,b5}''=0.25$\\
$k_{s,n2}'=0$ \ \ \ \ \ \ \ \ \ \ \  \ \hspace{0.1cm}  $k_{s,n2}''=0.05$\ \ \ 
\ \ \ \ \hspace{0.06cm} $k_{d,n2}=0.1$  \\
$k_{s,b2}'=0.002$ \  \ \ \ \  \ \ \ \hspace{-0.02cm} $k_{s,b2}''=0.05$ \\
$k_{d,b2}'=0.010$ \ \ \ \ \ \ \ \ \hspace{-0.03cm} $k_{d,b2}''=2$\ 
\ \ \ \ \ \ \ \ \ \ \hspace{0.07cm} $k_{d,b2}'''=0.05$ \\
$k_{s,c1}'=0.020$ \ \ \ \ \ \ \ \ \hspace{-0.03cm} $k_{s,c1}''=0.1$\
 \ \ \ \ \ \ \ \hspace{0.18cm} $k_{d1,c1}=0.01$ \ \ \ \ \  \ \ 
\hspace{-0.03cm} $k_{d2,c1}=0.3$\\
$k_{as,b2}=50$ \ \ \ \ \ \ \ \ \ \ \ \hspace{-0.07cm} $k_{as,b5}=50$ \ \ \
\ \ \ \ \ \ $k_{di,b2}=0.05$ \ \ \ \ \ \ \ \ $k_{di,b5}=0.05$ \\
$k_{s,20}'=0.005$\ \ \ \ \ \ \ \ \ \hspace{-0.02cm} 
$k_{s,20}''=0.06$\ \ \hspace{-0.08cm} \
\ \ \ \ \ $k_{d,20}=0.08$ \\
$k_{a,20}=1$\ \ \ \ \ \ \ \ \ \ \ \ \ \ \hspace{0.0cm}  $k_{i,20}'=0.1$ \ \
 \ \ \ \ \ \ \ \hspace{-0.1cm} $k_{i,20}''=10$\\
$k_{a,t1}'=0.04$ \ \ \ \ \ \ \ \ \ \ \ $k_{a,t1}''=2$ 
\hspace{-0.05cm} \ \ \ \ \ \ \ \hspace{0.3cm} $k_{i,t1}'=0$ \ \ \ \ \
\ \ \ \  \ \ \ \  $k_{i,t1}''=0.64$  \\
$k_{a,sbf}=1$ \ \ \ \ \ \ \ \ \ \ \ \ \hspace{0.05cm} $k_{a,mcm}=1$
\ \ \ \ \ \  \ \  $k_{a,swi}=1$
\ \ \ \ \ \ \ \ \  \ \hspace{-0.05cm}  $k_{i,mcm}=0.15$ \\
$k_{i,sbf}'=0.5$ \ \ \ \ \ \ \ \  \ \  \hspace{0.1cm} 
$k_{i,sbf}''=6$\ \ \hspace{0.05cm} 
\ \ \ \ \ \ \ \ $k_{i,swi}'=0.3$ \ \ \ \ \ \ \ \ $k_{i,swi}''=0.2$ \\
Characteristic concentrations(dimensionless)  \\ 
$[{\rm Cln3}]_{max}=0.02$ \ \ \ \ \ \ \ \ \ \ $[{\rm Bck2}]^0=0.0027$ 
\ \ \ \ \ \ \ \  \ \ \ \ \ \ $J_{d2,c1}=0.05$\\
$J_{a,sbf}=J_{i,sbf}=0.01$\ \ \ \ \ \ \ $J_{a,mcm}=J_{i,mcm}=1$
\\
$J_{a,swi}=J_{i,swi}=0.1$\ \ \ \ \ \ \ \ $J_{a,t1}=J_{i,t1}=0.05$\\
Kinase efficiencies(dimensionless)\\
$\epsilon_{c1,n3}=20$ \ \ \ \ \ \ \ \ \ \ $\epsilon_{c1,k2}=2$ \ \ \ \ \ \ \ \ 
$\epsilon_{c1,b2}=0.067$  \ \ \ \ \ \ \ \  $\epsilon_{c1,b5}=1$\\
$\epsilon_{i,t1,n2}=1$ \ \ \ \ \ \ \ \ \ \ $\epsilon_{i,t1,b2}=1$ \ \
\ \ \ \ \hspace{0.05cm}  $\epsilon_{i,t1,b5}=0.5$\\
$\epsilon_{sbf,n3}=75$ \ \ \ \ \ \ \ \ \ $\epsilon_{sbf,b5}=0.5$
\ \ \ \ \ \ $V_{i,20}=0.05$\\
Other parameters\\
$f=0.433$ \ \ \ \ \ \ \ \ \ \ \ \ \ $J_{n3}=6$ \ \ \ \ \ \  \ \ \ \ $D_{n3}=1$
\ \ \ \ \ \  \ \ \  \ \ \ \ $\mu=0.005776$
\\

\end{tabular}
\end{ruledtabular}
\end{table}
\newpage

}

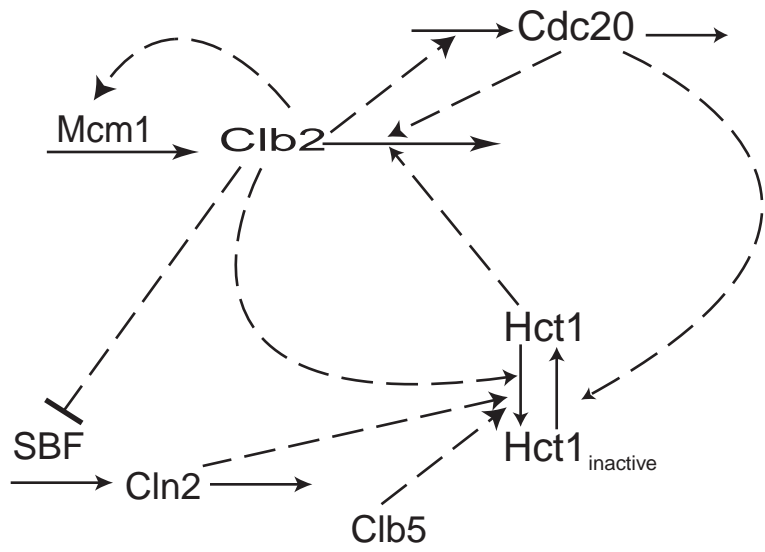
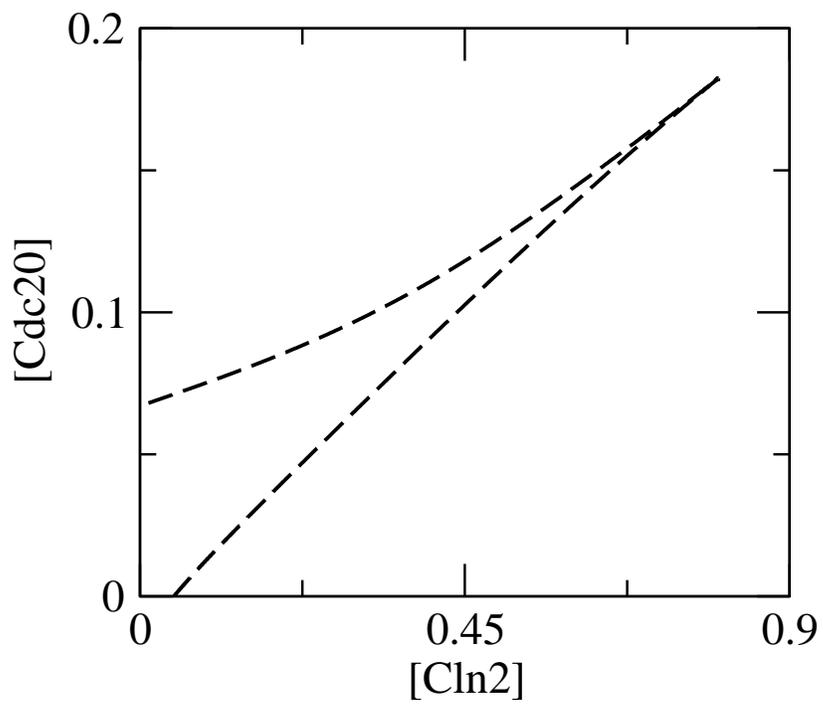
\end{document}